\documentclass[10pt, journal]{IEEEtran}

\usepackage{cite}
\ifCLASSINFOpdf
\else
\fi

\usepackage[cmex10]{amsmath}

\usepackage{algorithm}
\usepackage{algorithmic}
\usepackage[normalem]{ulem}

\usepackage{graphicx}
\usepackage[T1]{fontenc}
\begin{document}
%
% paper title
% can use linebreaks \\ within to get better formatting as desired

\title{BRVST: Efficient and Content-Expressive Information Matching Overlay in Wireless Networks}
%\title{BRVST: Efficient and Flexible Content Representation and Matching over Wireless Networks}

% author names and affiliations
% use a multiple column layout for up to three different
% affiliations

\author{\IEEEauthorblockN{Ying Li and Xin Wang\\}
\IEEEauthorblockA{Department of Electrical and Computer Engineering\\
Stony Brook University\\
Email: \{yingli, xwang\}@ece.sunysb.edu}
\thanks{\copyright 2014 IEEE. This paper has been accepted to MASS 2014. Personal use of this material is permitted. Permission from IEEE must be obtained for all other uses, including reprinting/republishing this material for
advertising or promotional purposes, collecting new collected works for resale or
redistribution to servers or lists, or reuse of any copyrighted component of this work in other
works.}}

%\author{\IEEEauthorblockN{Authors Name/s per 1st Affiliation (Author)}
%\IEEEauthorblockA{line 1 (of Affiliation): dept. name of organization\\
%line 2: name of organization, acronyms acceptable\\
%line 3: City, Country\\
%line 4: Email: name@xyz.com}
%\and
%\IEEEauthorblockN{Authors Name/s per 2nd Affiliation (Author)}
%\IEEEauthorblockA{line 1 (of Affiliation): dept. name of organization\\
%line 2: name of organization, acronyms acceptable\\
%line 3: City, Country\\
%line 4: Email: name@xyz.com}
%}

% make the title area
\maketitle

\begin{abstract}
%\boldmath
%\rev{Network information dissemination has revived to be an important and challenging research for mobile wireless networks since the prosperity of social network motivated smart phone applications and cloud-based services.}
Efficient and flexible information matching over wireless networks has become increasingly important and challenging with the popularity of smart devices and the growth of social-network-based applications. Some existing approaches designed for wired networks are not applicable to wireless networks, due to their overwhelming control overheads. In this paper, we propose a reliable and scalable binary range vector summary tree (BRVST) infrastructure for flexible information expression support, effective content matching and timely information dissemination over the dynamic wireless network. A novel attribute range vector structure has been introduced for efficient and accurate content representation and a summary tree structure to facilitate information aggregation. For robust and scalable operations over dynamic wireless network, the proposed overlay system exploits a virtual hierarchical geographic management framework. Extensive simulations demonstrate that BRVST has a significantly faster event matching speed, while incurs very low storage and traffic overhead, as compared with peer schemes tested.
\end{abstract}

\section{Introduction}
\label{Section:Introduction}

With the drastic growth of social and wireless application information data generated and consumed, it is emergent to establish a bridge infrastructure that can timely and accurately discovers and delivers the information to various parties of interests.

As an example of new era information service, a smart-phone user in a downtown block wants to obtain a recommendation for some restaurants while people close-by may be also searching for the same type of information. Another user just stepping out of a Thai cuisine is satisfied with the dining experience and would like to share this place with others. Other applications include traffic information posting and retrieval where users cooperatively contribute to and benefit from the real-time traffic reports.

These applications can be better met by a "contribute-and-benefit" pattern system. Publish/Subscribe (Pub/Sub) system is one of this type, in which subscribers specify their interests and publishers post advertisements. The system matches subscriptions with publications. Unlike client/server models, the Pub/Sub model decouples time, space, and flow between publishers and subscribers to provide flexibility in information distribution.
%which not only reduces resource consumption but also

Gryphon~\cite{gryphon} and SIENA~\cite{SIENA} were once popular Pub/Sub models in wire-line networks, however, their tree-based structure are not scalable in dynamic wireless network whose topology may constant change due to mobility and connection broken.
 %To address the attack: In the Introduction, the authors seem to claim that SIENA is not scalable.  However, SIENA's claims very much include scalability.  The authors suggest topologies change, but they do not explain what sorts of issues cause these changes (mobility, radio shadows, etc.).

 Many later attempts have been made to apply Pub/Sub infrastructure for wireless networks \cite{DRIP}\cite{GeoRendezvous}\cite{GeographicalContentPubSub}, where the information in the systems is roughly divided into several basic types. These platforms cannot efficiently support heterogeneous user application needs.

Different from conventional Pub/Sub systems which mainly categorize information into a few types for ease of implementation, the modern information system is expected to better meet the customized information needs of individual users. Besides the difference in categories, the heterogeneity of information is more generally resulted from different values or contents for the same type of information. In the restaurant recommendation example, the difference in the service time of a day or the average price level would totally distinguish restaurants and draw the interests of different groups of consumers, even when they provide the same type of foods. Simply ascribing information into coarse types (food, movie, car, etc.) cannot meet most application needs. On the other hand, completely expressing every detail of the information in words and matching over them is not feasible in reality. We need an information system that supports rich and accurate information content expression while efficiently reducing the representation complexity.
 %by the need to enhance information content expressiveness in a numerical sense beyond conventional simple type difference.}

In this paper, we propose a reliable and scalable content-expressive information matching and dissemination infrastructure in a large-scale mobile wireless network, which utilizes novel and efficient components as well as a location-based virtual management infrastructure for efficient storage, light-weight communications, and quick information match.

The main contributions of our work are:
\begin{itemize}
  \item
  %We propose a novel   While very few related literature has directly faced the challenges of the intangible variations and redundancies brought by the flexible representation of information in content Pub/Sub system as well as to allow a range value confinement to information.

   We propose a mechanism to flexibly and efficiently represent information with the combination of a set of elementary tuples for numerical expression of the content.

%  Our design realizes the true sense of content based Pub/Sub model to support unlimited different types of information with controllable accuracy of range confinements to the value by the free combination of a finite set of elements.
  \item We propose a novel Attribute Range Vector that allows flexible vector length adjustment based on the information accuracy requirement, and supports a unique simple bit-wise operation for quick content matching check, to facilitate accurate content representation as well as low-overhead in storage and transmission.

  \item We propose a Summary Tree structure to facilitate efficient aggregation of information, which significantly reduces the overhead for storing and transmitting information updates.

   %This data structure helps to significantly reduce the overhead for storing and transmitting information updates.

  \item Different from Pub/Sub systems, which generally match the publication over predefined or existing subscriptions, where subscribers usually have to wait, our scheme makes the matching process bidirectional so that all information can be promptly processed for matching.
\end{itemize}

The rest of the paper is organized as follows. In Section~\ref{Section:RelatedWork}, some related works are discussed. Section~\ref{Section:SystemOverview} gives the system basic conventions and an overview of the overlay structure. Section~\ref{Section:MainScheme} outlines the detailed design and algorithms for information matching of BRVST. %Section~\ref{Section:Mobility} discusses the node mobility and failures handling approaches.
Extensive simulations are evaluated in Section~\ref{Section:Simulation}. We conclude the work in Section~\ref{Section:Conclusion}.

\section{Related Work}
\label{Section:RelatedWork}
There are lots of studies on developing information matching mechanisms, among which Publish/Subscribe systems are once prevalent. However, the work on Pub/Sub systems over wireless networks is far less mature than that in wired networks.

Very few efforts have been made to support flexible content-based information matching and dissemination over wireless networks. One of the challenges is to accurately represent the content which often has a value range and to support efficient query on the ranges. R-Tree~\cite{R-Tree} supports range query for a single content attribute, but the structure consumes too large space when the information is composed of multiple attributes.
%As user interests are often heterogeneous and each interest is also composed of multiple attributes, it would consume too much space to maintain the infrastructure and incur high overhead to match each attribute.
Bloom Filter can also be modified to support range query. MDSBF~\cite{BloomfilterRange} combines multiple bloom filters with each one representing one attribute of the content. However, this can easily get into computational bottleneck as information volume increases, because the query on each attribute bloom filter requires several hashing operations.
% Unless modified to support valid aggregation, using several bloom filters to represent just one piece of information is unacceptable in performance for large loads.
TAMA~\cite{TAMA} has its own design to express numeric ranges. Its fixed granularity-level design, however, lacks the ability to balance between content representation accuracy and storage efficiency. Besides, TAMA maintains information in tables without aggregation, which is not efficient in both space and time complexity. Instead, our novel variable-length attribute range vector, which supports convenient aggregation, can not only flexibly represent numeric range of content to any desired accuracy level with low storage space, but also take advantage of simple bit-wise operations to facilitate efficient information matching.

Other types of systems such as \cite{TreePubSub} by Picco et al. assume tree-based topologies, which are hard to maintain and vulnerable to network topology changes. To avoid this drawback, the wireless network can be divided into regions for more efficient management and information distribution. DRIP~\cite{DRIP} groups nodes registered to different broker nodes into Voronoi regions whose shape and size could change over time. However, it may involve a high overhead to maintain the topology region especially over a mobile network. Based on virtual infrastructure, our design avoids the high overhead of region maintenance and also facilitates information aggregation to minimize information update changes.
%over dynamic wireless networks with constant node movement and information updates.

%Targeting for efficient information matching, our proposed BRVST provides a content-expressive platform that ensures high efficiency in storage, processing and communications, and is robust to network dynamics.

\section{Model Basics and System Overview}
\label{Section:SystemOverview}

%There are two major classes of publish/subscribe systems:
%(i) topic-based \cite{PubSubClassification} and (ii) content-based \cite{ContentPubSubModel}. In topic-based systems, subscribers join a group containing a topic of interest. Publications that belong to the topic are broadcasted to all members of the group. Therefore, publishers and subscribers must explicitly specify the group they wish to join. In content-based Pub/Sub systems, the matching of subscriptions and publications is based on content instead of being constrained to a few coarse groups.
%Therefore, these systems support more flexible and accurate representation of information. The main challenge in building such systems is to develop an efficient matching algorithm that can handle the flexibilities.

In this work, we adopt the notion of Publication and Subscription to distinguish information from the generators and to the consumers. The whole information space is built up with the basic element - attribute ($A_i, i=1,2,...$), which contains attribute name ($a_n$) specifying the identification of an attribute (numeric ID in realization), and attribute value ($a_v$) that specifies the content and is usually a numeric point or range. i.e. $A_i = \{a_n, a_v\}$.
%Although the number of supported attribute names in the system is finite, the permutation they create is far beyond enough for expressing the information space.
%For easy matching, in the system, each attribute is represented as an ID, and the attribute value range is within a pre-defined range limit.
% Each attribute is assigned a unique number as the Attribute ID, or AID, and is a shared knowledge of the whole system. For efficient processing, under agreement of common knowledge, any kind of attribute content, no matter strings or symbols, can always be numerated into values.

A subscription \textbf{s} is a conjunction of n attributes: \textbf{s} = $\{A_1 \wedge \cdots \wedge A_n\}$. %Table~\ref{Tab:Sub} shows an example subscription of sport tickets.
%\begin{table}[htbp]
%\centering
%\caption{An example Subscription of sports ticket}
%\begin{tabular}{|c|c|} \hline
%Attribute Name&Attribute Value\\ \hline \hline
%Price & 100 $\sim$ 300\\ \hline
%Seat Rank & 1 $\sim$ 3\\ \hline
%Date & 20120701 $\sim$ 20120930\\ \hline
%%\Where & * NY\\ \hline
%\end{tabular}
%\label{Tab:Sub}
%\end{table}
A publication \textbf{p} is a disjunction of attributes: \textbf{p} = $\{A_1 \vee \cdots \vee A_n\}$, and is also referred to as an event. Conventionally the attribute value of a subscription could either be a numeric point or a range, while that for publication is assumed only to be a numeric point, and many literature studies \cite{TAMA}\cite{ContentPubSubModel} have followed this convention. However, very often some attributes of the information, when generated, are not absolute point values. For example, the video surveillance data could have its \textit{time} attribute as a range which confines the start and end points of a video segment. So our design also supports range value for a publication attribute.

We assume all data published are trustful, and there is no fraud or spam. Detecting malicious data is not our focus.

For users to get more precise information, we consider a publication and a subscription to match each other iff: for each attribute existing in the subscription, the same attribute must also exist in the publication; and for the common attributes, those from the publications must have their value ranges contained by the value ranges of the corresponding attributes in the subscription. i.e.$\forall A^s\in s, \exists A^p\in p: (a_n^p = a_n^s, a_v^p \subseteq a_v^s)$,
where the superscript $^s$ denotes the subscription, while $^p$ denotes the corresponding terms for a publication.

% system overview starts here-------------------------------------

\begin{figure}[h]
\centering
\includegraphics[width=3.5 in]{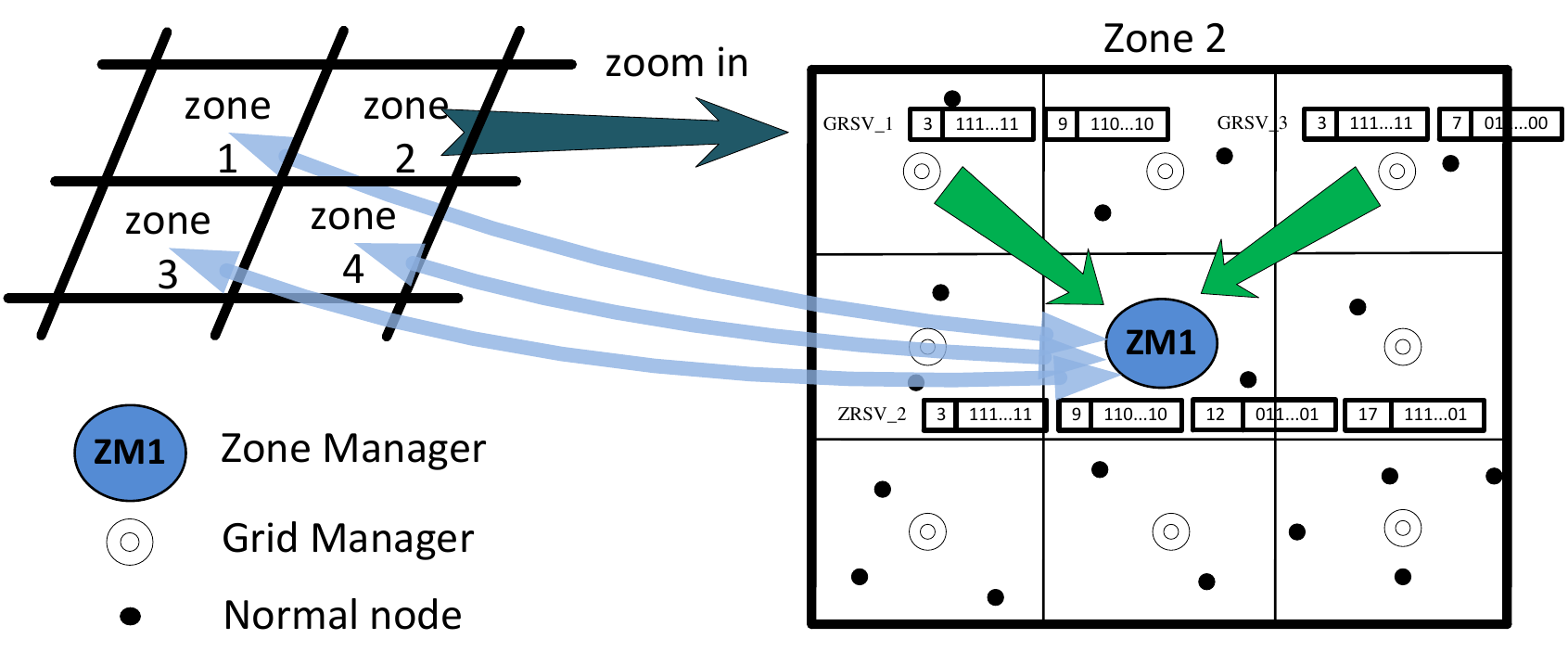}
% where an .eps filename suffix will be assumed under latex,
% and a .pdf suffix will be assumed for pdflatex; or what has been declared
% via \DeclareGraphicsExtensions.
\caption{An example system where each zone has 9 grids. The zone manager collects Pub/Sub messages from grids within the zone and aggregates them into control messages to exchange with other zones.}
\label{fig:GridSystem}
%\vspace{-0.2in}
\end{figure}

In order to make the infrastructure scalable and more robust to the network dynamics, we introduce a virtual management infrastructure where the network space is mapped into virtual zones each consisting of a set of virtual grids (Fig.~\ref{fig:GridSystem}). With many wireless devices equipped with GPS receivers or having other methods of localization~\cite{GeoRendezvous}, the grid and zone which a node belongs to can be easily calculated based on node location in reference to a reference virtual origin and a pre-determined grid or zone size~\cite{multicastTOC}. There is no need of a complicated scheme to create and maintain the virtual grids or zones. The grid size can be determined by the system based on the application scenarios and performance tradeoffs. Its effects is studied in Section~\ref{Section:Simulation}. %Fig.~\ref{fig:MobilityAndGridSize}-(b).

%Composed by grids, the zone has the similar influence of its size.}
%\rev{The influence of grid and zone size is examined in Section~\ref{Section:Simulation}.}

Each grid can elect a Grid Manager (GM) for Pub/Sub message collection, aggregation and matching within the grid. Each zone also has a Zone Manager (ZM) responsible for Pub/Sub aggregation, matching, data catching over grids within the zone. The schemes for leader election and maintenance have been proposed by many literature work~\cite{multicastTOC} which can be leveraged in our system
%, and the election can take into account factors such as the power and resources of the nodes as well as the node distance to the center of the grid or zone
. The managers can be static or mobile, depending on the system application scenarios.
%More on GM and ZM maintenance is discussed in Section~\ref{Section:Mobility}.
%\rev{It is worth noting that our system works extremely efficient and reliable over complete mobile Ad-hoc environment, the ZM or GM could though be made static or backbone connected, which only makes the life easier.}

%In order to better manage the network and o avoid the frequent topology state changes resulting in node mobility,
 %we introduce a zone-based scheme allowing the whole network to be virtually divided by rectangular grids denoted by unique grid IDs (GID). Different from conventional cluster structures, there is no need to involve a complicated scheme to create and maintain the zones. Since each node is aware of which zone it belongs to based on its position coordinates (i.e. by means of GPS components) and the reference coordinates of region boundaries known to the whole network. In each grid, there will be a local manager, referred to as the grid manager(GM), which is responsible for subscription/publication collection and aggregation. This grid manager is elected with respect to the power and resources it possesses from all the mobile nodes within this grid.

%The whole system is hierarchically managed as in the example shown in Figure~\ref{fig:GridSystem}. Several grids form a bigger area referred to as a zone, which is also assigned a unique ID called the ZID. A powerful node geographically close to the center of the grid is elected to be the Zone Manager (ZM), which is responsible for subscription and publication aggregation, matching, data catching, etc. In case of zone manager moving out of its confined range(usually centered at but smaller than the scope of a zone), another node within the confined range will be elected as the new zone manager and take over charges. There are several mature solutions for electing nodes as manager of a network area regarding the resources they possess \cite{multicastTOC}, which is beyond the focus of this paper. The function of both GM and ZM will be discussed in detail in next section.

Event matching and Pub/Sub message update are both performed on demand. Subscriptions and publications in a grid are collected and aggregated. Although nodes may frequently move in and out of a grid, the aggregate filter may stay unchanged. Messages are sent to the upper level ZM only upon the change of aggregate filter. This will significantly reduce the overhead for Pub/Sub message transmission and matching in a dynamic wireless network. A ZM maintains the Pub and Sub information of the grids within its zone with efficient data structures to be introduced in Section~\ref{Section:MainScheme}, and the Pub/Sub information of the whole zone can be similarly further aggregated. As many mobile users have interests in close-by information, the aggregate filters only need to be shared among nearby zones or zones identified with Pub/Sub relationship.

%Depending on the publication types and the relative number of subscribers and publishers, the system could support more active zone-level broadcast of aggregate publication or subscription messages upon need. As our goal is to address the challenge of supporting efficient content-based information dissemination, the study of optimal global-level message distributions is not our focus.

Any new subscription or publication will trigger the event matching process within its own zone first, then matching at other zones whose aggregate filters imply potential chance of match will initiate. This will significantly reduce the data matching and distribution overhead. Once a publication is matched with one or more subscribers, the overlay structure will then deliver the data to these destinations using the stateless geographic multicasting, RSGM~\cite{multicastTOC}, for reliable and low overhead transmissions. The detailed routing process is beyond the scope of this paper.
%By completing each Pub/Sub matching and delivering between the data generator and the consumer, our system accomplishes the information distribution.

\section{Bidirectional Content Matching}
\label{Section:MainScheme}
%A good information service system should strive to bring the most precise information to the interested requesters with low delay. In addition, our infrastructure also enables flexible information access--either matching new event with existing subscriptions or allowing new subscriber to timely retrieve information from existing publications.
%In addition, our Pub/Sub infrastructure will also exploit Published information to enable more efficient and timely information delivery.

In many conventional Pub/Sub systems, the subscriptions are specified before the publications.
%either at the system initialization time or collected before events happen.
However, some subscribers may indicate their interests on some data that have been published before. Simply throwing away the published data when they cannot match the current subscriptions would waste the system resources consumed for the information matching and distribution.
% dissatisfy the information need of the customers.
%For example, a scientist may be interested in the environmental data published earlier, and a police officer may want to retrieve past monitoring data upon detecting an event.
%Or, if the published data are simply ignored once they cannot match the current subscriptions, it would not only waste the system resources and transmissions that have been made for this matching attempt but also dissatisfy the information need of the customers.
%the communication pattern goes more often in the opposite direction, where a user-triggered application would need a specific type of data to accomplish its task, then it tells the system about its needs, and waits for the data to be found and delivered back.
Instead, if the publications can be stored even if they did not find match, the system could immediately deliver the data to later subscribers once they have interests.

%Besides basic Pub/Sub requirements, in a mobile information system with scarce wireless bandwidth, unreliable and less powerful wireless devices, and possible constant node movement, the challenge is big in updating the Pub/Sub states timely to ensure reliable data delivery to the requested subscribers.

%The bandwidth of wireless medium is scarce, and wireless devices prefer lower processing due to the limitation of their computational power and need of energy conservation.

In this work, we propose an efficient bi-directional content matching infrastructure, so that newly published data will be timely distributed to existing subscribers matched and new subscriptions can also trigger the retrieving of interested data already published quickly. In face of the challenge of representing the rich contents while not significantly sacrificing system performance, we novelly propose simple binary bit vectors and summary tree structure to facilitate flexible content-expressive information matching and dissemination processes at low overhead for storage, transmission and computation.
%We exploit virtual grids to ensure scalable, light-weight and robust information management in a dynamic mobile wireless environment.
%\rev{ and our performance studies in Section~\ref{Section:Simulation} demonstrate that the system is .}
% in a dynamic mobile wireless environment, and

\subsection{\textbf{Binary Vector and its Operations}}
\label{vector}
Content-based information system can potentially support flexible user information need, but at the same time poses high challenges for information representation and matching. We introduce simple  {\em Attribute Range Vector} to facilitate light-weight content-expressive management while not compromising the accuracy of information matching.

%Mobile devices are often energy-constrained and require lower processing overhead, and wireless bandwidth is limited. There is a need to reduce the control overhead and to avoid transmitting unwanted data due to inaccurate information matching. To address these challenges, we introduce simple  {\em Attribute Range Vector} to facilitate light-weight content-expressive management while not compromising the accuracy of information matching.

\subsubsection{\textbf{Attribute Range Vector (ARV)}}
\label{FalsePositiveProof}
We propose a binary bit vector named Attribute Range Vector (ARV) to flexibly represent the numeric range values of an attribute, referred as the target range. The target range could be a single point value as well. An ARV has a small size and is easy to process. The numeric value of an attribute is generally limited within predefined boundaries, which can be determined in advance by the system based on some common knowledge. For example, the temperature of the weather has an lower and upper limit in physical world. A subscriber could indicate her interest by setting a target range within the limit defined by the system. To facilitate flexible range matching, the predefined limit range is divided into $N$ smaller equal segments, while the value of $N$ can vary based on the matching accuracy requirement. An $N$-bit ARV is formed by representing whether a segment matches a content range, following the steps below:

\begin{description}
  \item[\textbf{Step0}:] Set the initial segment to be the whole predefined limit range.
  \item[\textbf{Step1}:] Check if the target attribute value range falls into some existing segments with each occupied more than $\alpha$ (percentage) of the segment range, an accuracy threshold desired. If so, goes to the next step; otherwise divide each of the current segments into equal halves, and continue this step.
  \item[\textbf{Step2}:]Make an $N$-bit vector with $N$ equal to the current number of segments, with each bit indicating if the attribute range overlaps the corresponding segment range, 1 yes, and 0 no.
\end{description}

From the above ARV construction process, we can see that the number of bits of the vector can only be the power of 2, i.e., $N=2^{i}$, $i=0,1,2,3$..., and the length of ARV can be continuously doubled until a desired representation accuracy is achieved. The threshold $\alpha$ trades off between accuracy and simplicity of the message representation.
%along with its impact on message volume consumptions in the simulation.}

%\vspace{-0.12in}
\begin{figure}[h]
\centering
\includegraphics[width=2.2in]{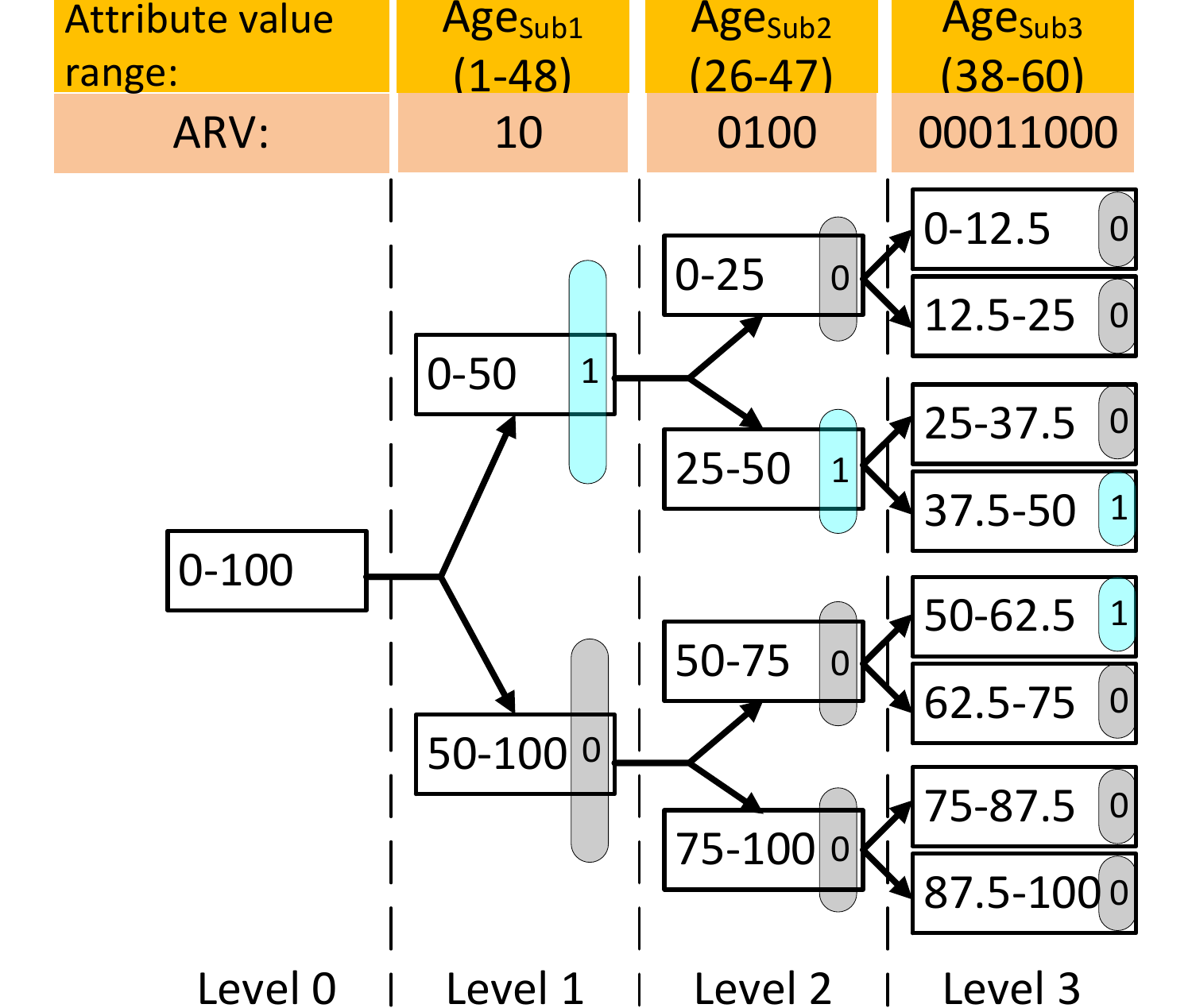}
% where an .eps filename suffix will be assumed under latex,
% and a .pdf suffix will be assumed for pdflatex; or what has been declared
% via \DeclareGraphicsExtensions.
\caption{The segment division procedure in constructing an ARV.}
\label{fig:BinarySplitTree}
%\vspace{-0.14in}
\end{figure}

For example, the attribute \textit{Age}, often involved in social network applications, is limited within 0 to 100. Three subscriptions that contain the attribute \textit{Age} are: $Age_{Sub1}$ 1-48, $Age_{Sub2}$ 26-47, and $Age_{Sub3}$ 38-60. Their corresponding ARVs are obtained by constructing a split tree following the above steps as shown in Figure~\ref{fig:BinarySplitTree}, with the level $i$ having $2^{i}$ segments. Suppose the threshold $\alpha$ is set to 90$\%$ in this example. $Age_{Sub1}$ falls into the segment 0-50 and the fitting ratio of the target range 1-48 is $48/50$, which is larger than the threshold $\alpha = 90\%$. So this segment is accurate enough to represent the target range and the ARV for $Age_{Sub1}$ is 10. $Age_{Sub2}$ apparently falls into the 0-50 segment of level 1, however, this range is not very accurate. We further divide the overall range into 4 new segments at the level 2, so the range 26-47 falls into the segment 25-50. We can use 4-bit vector 0100 to represent this 4-segment coverage, with the left most bit standing for the segment of the lowest value. The target range 38-60 of $Age_{Sub3}$ spans across the 0-50 segment and the 50-100 segment at the first-level of the split tree, but these two segments are inaccurate in representing the target range. If we go deeper into the level 3, the segment 37.5-50 $\&$ 50-62.5 will be accurate enough with the resulting ARV 00011000.

A shorter ARV is always preferable to reduce the transmission and storage overhead. The ARV bit vector is checked after each modification for the potential of simplification. Except level 0, the number of bits in an ARV is always even and in the power of 2. When the length of ARV is larger than 1, starting from one side of the vector, if {\em every} consecutive 2-bit has the same value (both '1' or both '0'), the length of the vector can be reduced into half by taking every other bit to form a new ARV. For example, 1100 can be reduced to 10, but not 0110 nor 0111 which does not have the same value for consecutive 2-bit. The simplification operation will continue without losing the accuracy of the information until the vector cannot be further simplified.
%This operation guarantees the saving of storage without losing accuracy of the information.

Likewise, a given vector could also be extended by $2^{i}$ ($i$=1,2,3...) times when needed by simply doubling the bit patterns. This feature is extremely useful in the matching process we will discuss later, where two or more ARVs need to be adjusted to have the equal length before they can be compared or merged.

The proposed ARV is the elementary component of Subs and Pubs, and some other aggregated management structures at different hierarchical levels are composed of ARVs.

Under our matching rule specified in Section~\ref{Section:SystemOverview}, the attribute of a publication is considered to match with that of a subscription when its value range is contained by that of the same attribute of the subscription. When the same attribute for Pub and for Sub are represented with 2 ARVs respectively, and are scaled to the same length of bits, the matching rule now translates into: the bit positions where the Pub ARV has '1' need also to be '1' for the Sub ARV. There will be a slight chance for two ARVs satisfying this criteria to be actually not matched due to improper choice for the afore mentioned accuracy level threshold $\alpha$. As these attribute-level false positives get significantly large, an event-level false positive will happen which causes erroneous matching result and thus unwanted traffic overhead. The impact of threshold $\alpha$ on the event-level false positive rate is studied in Section~\ref{Section:Simulation}.

\subsubsection{\textbf{ARV Merge}}

A merge operation is needed for information aggregation. As the length of the vectors could only be the power of 2, two vectors of different lengths can always be made equal by doubling the length of the shorter one several times as previously mentioned.
%by extending the shorter one by $n$ times with $n$ being the ratio of the long length over the short length and is in the power of 2. Each bit of the shorter ARV can be simply repeated $n$ times without changing the meaning as introduced above.
Suppose we want to merge the same attribute vector of Sub1 $\&$ Sub2 which are 0100 $\&$ 10 respectively, we only need to scale up 10 by repeating each of its bit once to get 1100, and the merge can be completed by only a simple bitwise "OR" between 0100 $\&$ 1100 to get the result 1100. This number can in turn be simplified into 10 without losing the accuracy. This indicates that the segment 0-50 can represent the merge of the ranges 26-47 and 1-48. As the accuracy level for each segment is ensured to be higher than $\alpha$, the accuracy of the ARV will not be impacted when it is scaled up or down. The merge operation is always carried at the length of longest ARV thus over the finest level of segments, and the merge of ARV will maintain the accuracy level.
{\em The ARV's merit for convenient merge operation is critical to information aggregation which contributes to very low storage and transmission overhead.}

\subsubsection{\textbf{Match of ARVs}}
\label{Section:ARVmatching}
Our purpose of introducing ARV is to facilitate fast information matching, which could be easily achieved with fast bit-wise operations under the following conditions:

A subscription, represented by conjunctions of attributes like A$\wedge$B$\wedge$C, where A, B and C are three different attributes, is considered to be matched only if all the attributes are satisfied. A publication is allowed to have additional attributes than A$\vee$B$\vee$C, i.e. A$\vee$B$\vee$C$\vee$F$\vee$G, to still be considered as matching the subscription, as long as all the attributes of the subscription (A, B and C in this example) are satisfied on their values. This convention intuitively means that subscribers will always accept information that is more elaborate than their expectations.

%\begin{itemize}
%  \item A subscription, represented by conjunctions of attributes like A$\wedge$B$\wedge$C, where A, B and C are three different attributes, is considered to be matched only if all the attributes are satisfied. A publication is allowed to have additional attributes than A$\vee$B$\vee$C, i.e. A$\vee$B$\vee$C$\vee$F$\vee$G, to still be considered as matching the subscription, as long as all the attributes of the subscription (A, B and C in this example) are satisfied on their values. This convention intuitively means that subscribers will always accept information that is more elaborate than their expectations.
%  \item Although the publication attributes could also have range values, they are relatively small and condensed compared to those of a subscription. Publication attributes could not have completely separable ranges, as otherwise the publication can always be divided into separate publications.
%\end{itemize}

\begin{figure}[h]
\centering
\includegraphics[width=3 in]{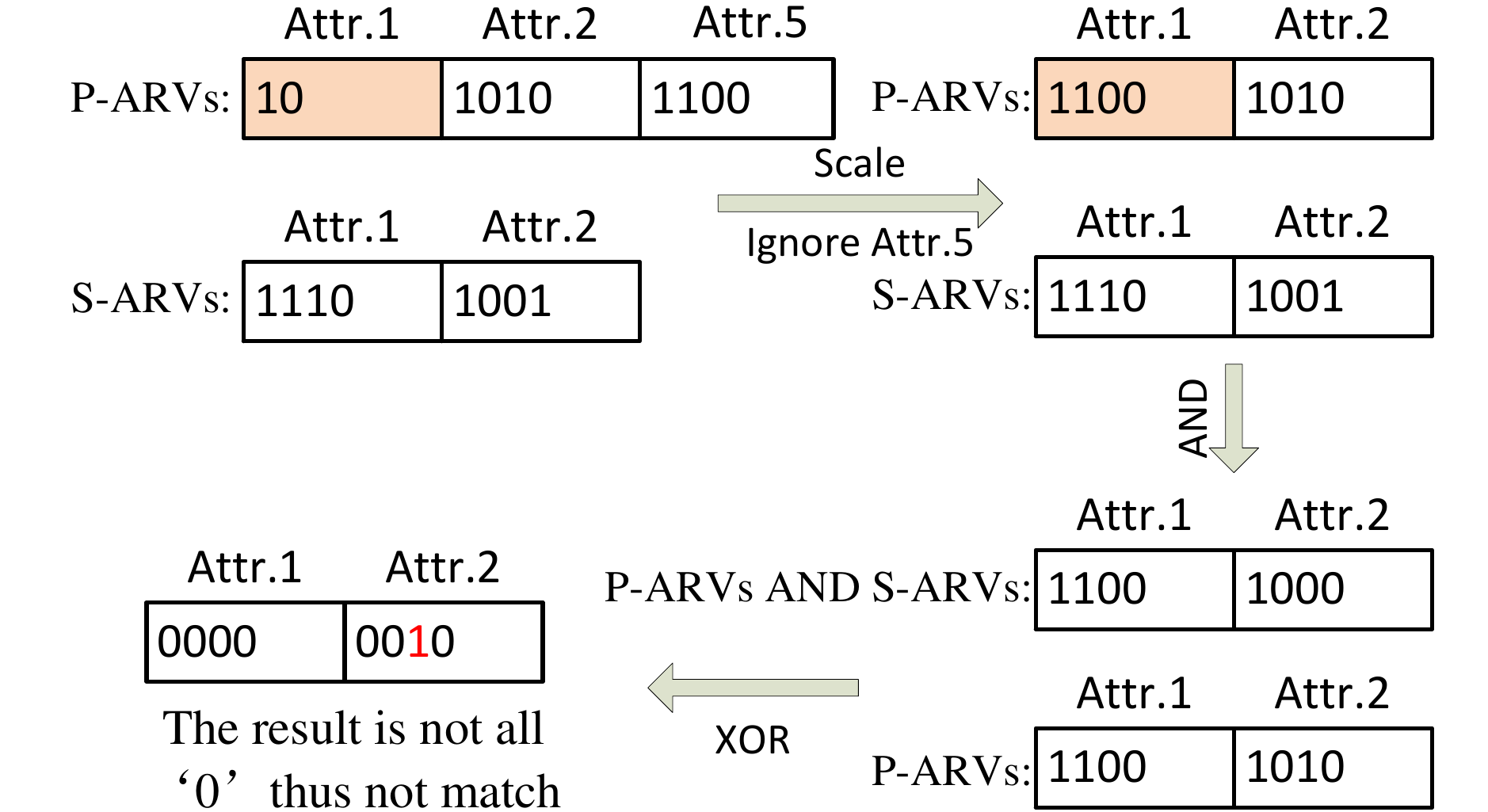}
% where an .eps filename suffix will be assumed under latex,
% and a .pdf suffix will be assumed for pdflatex; or what has been declared
% via \DeclareGraphicsExtensions.
\caption{The bit-wise matching evaluation of a Pub and Sub.}
\label{fig:BitVectorMatchOperation}
%\vspace{-0.18in}
\end{figure}

For differentiation and ease of referral, an subscription and publication attribute range vector are called respectively an S-ARV and P-ARV. If one or more attributes of the subscription are not included by the publication, we can immediately claim they do not match each other, given the conditions above. Otherwise they are further checked. First all the S-ARVs and P-ARVs are respectively concatenated following the corresponding order as shown in Figure~\ref{fig:BitVectorMatchOperation}, with all the redundant P-ARVs ignored and each corresponding pair of P-ARV and S-ARV scaled to the same length. Then the Sub and Pub are considered to match each other if and only if all bits after the following operations are 0: The cascaded P-ARVs vector and S-ARVs vector first have the bitwise AND operation, and the result XOR with the original cascaded P-ARVs vector.

%The first operation is applied to find the matched attribute ranges of the publication and subscription, and the second operation ensures that all the attributes of the publication fall into the ranges of the subscription.
%\vspace{-0.15in}

Figure~\ref{fig:BitVectorMatchOperation} gives an example. The subscription has 2 attributes, and the publication has 3 attributes. To perform the matching, the attribute 1 (Attr.1) of the publication is scaled to 4 bits, while the Attr.5 is omitted because it is not involved in the subscription. Then bitwise operations are carried out: (P-ARVs \texttt{AND} S-ARVs) \texttt{XOR} P-ARVs, and the result is not all '0' thus is not a match. Because Attr.2 of the publication has a '1' in the bit position where the subscription Attr.2 does not, which means the attribute 2 value range of the publication is out of that of the subscription.

%\textit{\textbf{Theorem 2}: The (P-ARVs \texttt{AND} S-ARVs) \texttt{XOR} P-ARVs operations give correct matching result.}
%
%\textit{\textbf{proof}:} We prove this theorem by contradiction. Suppose after the two steps of operations, all the bits of the remaining vector are '0's, but the Pub and Sub are actually not matched.
%%Since they are not matched,
%Before the two-step operations when the original P-ARVs and S-ARVs are cascaded in order and scaled to the same length with redundant P-ARVs omitted, there must be some bit positions that are '1's in the cascaded P-ARVs vector are '0's in the same positions of the cascaded S-ARVs vector, according to the matching principle. So if we take P-ARVs vector \texttt{AND} S-ARVs vector, the resulting vector must have all '0's in the aforementioned bit positions. Then if this vector \texttt{XOR} with the P-ARVs vector which has all '1's for those specific bits, those specific bits would all be '1's, which contradicts with the assumption of having all '0's in the resulting vector. The argument is proved.
\subsection{\textbf{Basic System Architecture and Maintenance}}

As introduced in Section~\ref{Section:SystemOverview}, we take the virtual grid as the lowest-level management unit for a multi-hop wireless network. Each grid has a Grid Manager (GM), and a set of grids form a zone that is under the control of a Zone Manager (ZM). The example system in Figure~\ref{fig:GridSystem} is split into multiple zones with each zone being composed of 9 grids. In this section, we present the functions at each level of our infrastructures.

\subsubsection{\textbf{Subscription Maintenance at the Grid Manager}}
A subscriber sends its subscription to its grid manager on demand. Each subscription message is a concatenation of all its attributes, i.e. all the corresponding ARVs. There are possibly many subscriptions in an information-dense area. Simply storing and transmitting all subscriptions would not only incur a large overhead in traffic and storage but also difficult to track the frequent subscription changes due to the user mobility and frequent user interest changes. On the other hand, selectively ignoring some of the subscriptions would compromise the system performance. In our system, the GM will aggregate the subscriptions by finding the minimum representative subscription set to represent all the subscriptions within the grid before recording them and sending them to the upper level.

%%\vspace{-0.15in}
%\begin{figure}[h]
%\centering
%\includegraphics[width=3.5in]{SubExample.pdf}
%% where an .eps filename suffix will be assumed under latex,
%% and a .pdf suffix will be assumed for pdflatex; or what has been declared
%% via \DeclareGraphicsExtensions.
%\caption{Example of a subscription request.}
%\label{fig:SubExample}
%%\vspace{-0.15in}
%\end{figure}

%Two subscriptions could share some common attributes, and the attribute range of a subscription could contain the attribute range of another subscription. In the second case, we can take the attribute range that is contained by many other subscriptions as the summary attribute range. A summary subscription is the one that contains a set of summary ranges of most common attributes.

Two subscriptions could share some common attributes, and the attribute set of a subscription could contain all the attributes of another subscription. In the second case, if the value ranges of the common attributes overlap each other to some extent, we could take the subscription which has all its attributes contained by the other subscription as the representative subscription of both subscriptions. However, if the value ranges of the common attributes do not have any intersection, then using one subscription to represent the other is not appropriate. We use an example to illustrate this aggregation principle. Suppose there are 2 subscriptions in a grid, SUB1: A and SUB2: A$\wedge$B$\wedge$C, where A, B and C are different attributes. According to our scheme, since all publications that contain the attribute A including the ones that also contain B and/or C will all be routed to this grid for further matching, thus taking SUB1 as the representative subscription, compared to otherwise having both SUB1 and SUB2, will help reduce the subscription information storage and control traffic without sacrificing the completeness of subscription information in this grid. Once receiving the information based on the aggregate filter, the GM will further match the information with individual subscription to determine if the information matches all the criteria of a subscriber. Thus aggregation reduces the message and data transmission between the ZM and GM, but does not sacrifice the accuracy requirement of each subscriber.
%\vspace{-0.11in}
\begin{figure}[h]
\centering
\includegraphics[width=3in]{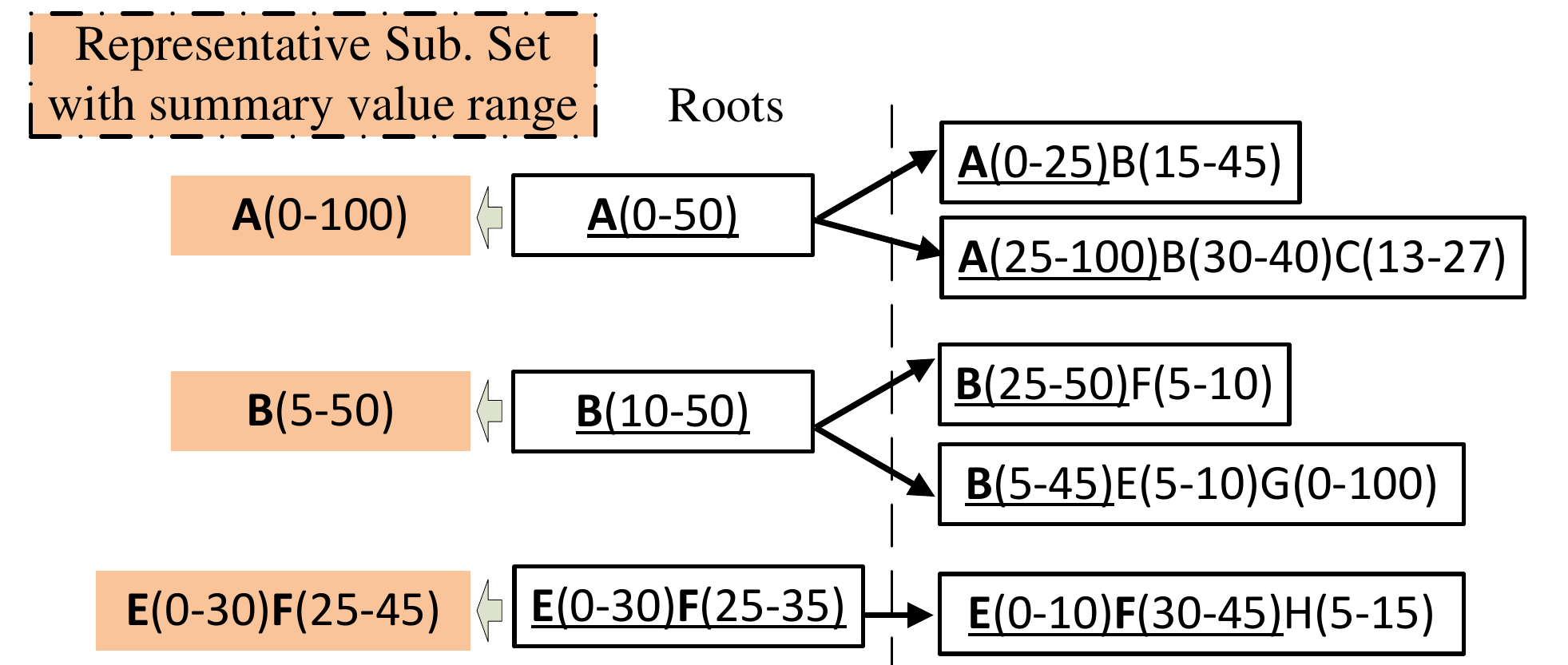}
% where an .eps filename suffix will be assumed under latex,
% and a .pdf suffix will be assumed for pdflatex; or what has been declared
% via \DeclareGraphicsExtensions.
\caption{Summary forest with attribute summary value range in shade.}
\label{fig:summaryTree}
%\vspace{-0.14in}
\end{figure}
The subscription aggregating process can be realized through a summary tree, which is actually a forest containing several separate trees as shown horizontally in Figure~\ref{fig:summaryTree}. All the subscriptions of a tree will be represented by its root, and a tree node will contain all attributes of the root. There is also a summary range attached to each root shown as the shaded block in Figure~\ref{fig:summaryTree}, obtained by merging ('OR' operation) the value range of common attributes (underscored in Figure~\ref{fig:summaryTree}) of all the subscriptions on a tree.  When determining if a node should be inserted into a tree, we will check if some of its attributes are the same as the root and if the attribute ranges overlap the current summary ranges. The summary ranges of all trees form the representative subscription set of the grid as shown on the left side of the dash line in Figure~\ref{fig:summaryTree}.

Algorithm~\ref{Alg:add2summarytree} shows how to add a subscription into the current summary forest. On lines 3-10, a new subscription will become either the child or the parent of an existing root, depending on whether it contains all the attributes of a root or all of its attributes are contained by a root of the forest, with the value ranges of corresponding common attributes overlapping each others. Otherwise, the subscription will be made a new stand alone root, as shown on lines 12 and 16. On line 18, after inserting the new subscription, the summary value range attached to the root of the affected tree will be updated. Line 19 checks whether trees can be merged to one another to reduce the number of trees in the forest, i.e., the size of the forest, every time the summary value range of a tree is changed, by examining whether one tree root can be inserted as the child of another tree root following the similar criteria.
%\note{The concept of summary set is very confusing here. It is not clear it represents all nodes on a tree, or all root nodes. Please check the correctness of the places.}
%\vspace{-0.1in}
\begin{algorithm}
\caption{Adding a subscription $s$ into the summary forest}
\label{Alg:add2summarytree}
\begin{algorithmic}[1]
\begin{small}
\IF {there are already nodes in the forest}
\FOR {each root node $R_i$ of the forest}
\IF {the subscription $s$ contains all the attributes in $R_i$}
\IF {the summary value range of each attribute in $R_i$ overlaps that of $s$}
\STATE insert $s$ as the child of $R_i$ into the summary tree;
\ENDIF
\ELSIF{$R_i$ contains all the attributes of $s$}
\IF {each attribute value range of $s$ overlaps the summary value range of the same attribute in $R_i$}
\STATE make $s$ the parent of $R_i$ as the new root;
\ENDIF
\ELSE
\STATE make $s$ a new root of the forest;
\ENDIF
\ENDFOR
\ELSE
\STATE make $s$ a new root of the forest;
\ENDIF
\STATE Adjust the summary value range of the affected tree.
\STATE Check whether the forest can be reduced by merging trees.
\end{small}
\end{algorithmic}
\end{algorithm}

%\vspace{-0.1in}

For illustration, suppose a grid has the following subscriptions with letters representing different attribute names: A(0-50), B(10-50), A(0-25)B(15-45), A(25-100)B(30-40)C(13-27), B(25-50)F(5-10), E(0-30)F(25-35), B(5-45)E(5-10)G(0-100), E(0-10)F(30-45)H(5-15). Applying them one after another with Algorithm~\ref{Alg:add2summarytree} will generate a summary forest as shown in Figure~\ref{fig:summaryTree}.

%%\vspace{-0.1in}
\begin{algorithm}
\caption{Removing a subscription $s$ from the forest}
\label{Alg:deletesummarytree}
\begin{algorithmic}[1]
\begin{small}
\IF{$s$ is a root of the forest}
\STATE delete the tree originated from root $s$;
\FOR {each children node of $s$}
\STATE apply Algorithm~\ref{Alg:add2summarytree};
\ENDFOR
\ELSE
\STATE delete $s$ from the summary tree;
\ENDIF
\STATE Adjust the summary value range for each affected tree.
\STATE Check whether the forest can be reduced by merging trees.
\end{small}
\end{algorithmic}
%%\vspace{-0.1in}
\end{algorithm}

Algorithm~\ref{Alg:deletesummarytree} works to remove a node in response to unsubscription. On lines 1-5, if the subscription to be deleted is the root of a tree, then this whole tree is removed with all the non-root nodes reinserted into the forest by applying algorithm~\ref{Alg:add2summarytree} one by one. If the subscription is not a root, it is simply deleted from the tree as shown on lines 6-7. Then the affected trees will have their summary value ranges updated accordingly on line 9. Line 10 works similarly as the last line of Algorithm~\ref{Alg:add2summarytree} to reduce the forest size.

%Each time after the range summary has changed, Algorithm~\ref{Alg:forestmerge} will be applied to reduce the forest. The algorithm works
%
%\begin{algorithm}
%\caption{Forest reduction after each range summary action}
%\label{Alg:forestmerge}
%\begin{algorithmic}[1]
%\IF{$s$ is a root of the forest}
%\STATE delete the tree originated from root $s$;
%\FOR {each children node of $s$}
%\STATE apply Algorithm~\ref{Alg:add2summarytree};
%\ENDFOR
%\ELSE
%\STATE delete $s$ from the summary tree;
%\ENDIF
%\STATE Adjust the merged summary range for each affected tree, then apply Algorithm~\ref{Alg:forestmerge}.
%\end{algorithmic}
%\end{algorithm}

Each GM will maintain a subscription summary forest, and updates the trees in response to the changes of subscription from individual subscribers within the grid. When a node wants to send a new subscription, modify or unsubscribe its existing subscription, it will send a message with the affected sub through on-demand light-weight geographic routing~\cite{SOGR} to the GM. The GM will either insert or delete the subscription following the Algorithm~\ref{Alg:add2summarytree} or~\ref{Alg:deletesummarytree}. A new action may change the representative set. In many cases, however, {\em individual subscription changes will not lead to the change of the aggregated information summary at the root level of the tree}. This feature is very important. It helps to increase the stableness and significantly reduce the information maintenance overhead in a wireless environment with possible constant node movement and thus frequent subscription changes. The representative set is forged into a vector, named Grid Representative Set Vector (GRSV) as shown in Figure~\ref{fig:SOF2ZRSV} by cascading each subscription from the representative set. The GRSV will be sent to the ZM upon its change to reduce the update overhead.
%To further protect against failure, ZM can send some messages to the subscribing grids within its zone if there are no information dissemination for a pre-defined long period. On the other hand, GM will send updated GRSV to ZM if it does not receive any information after a predefined period or when the grid receives unwanted traffic from ZM. ZM or GM can be re-elected upon failure using scheme similar to that proposed in}~\cite{multicastTOC}.

\subsubsection{\textbf{Subscription Maintenance at the Zone Manager}}

%\vspace{-0.1in}
\begin{figure}[h]
\centering
\includegraphics[width=3.5in]{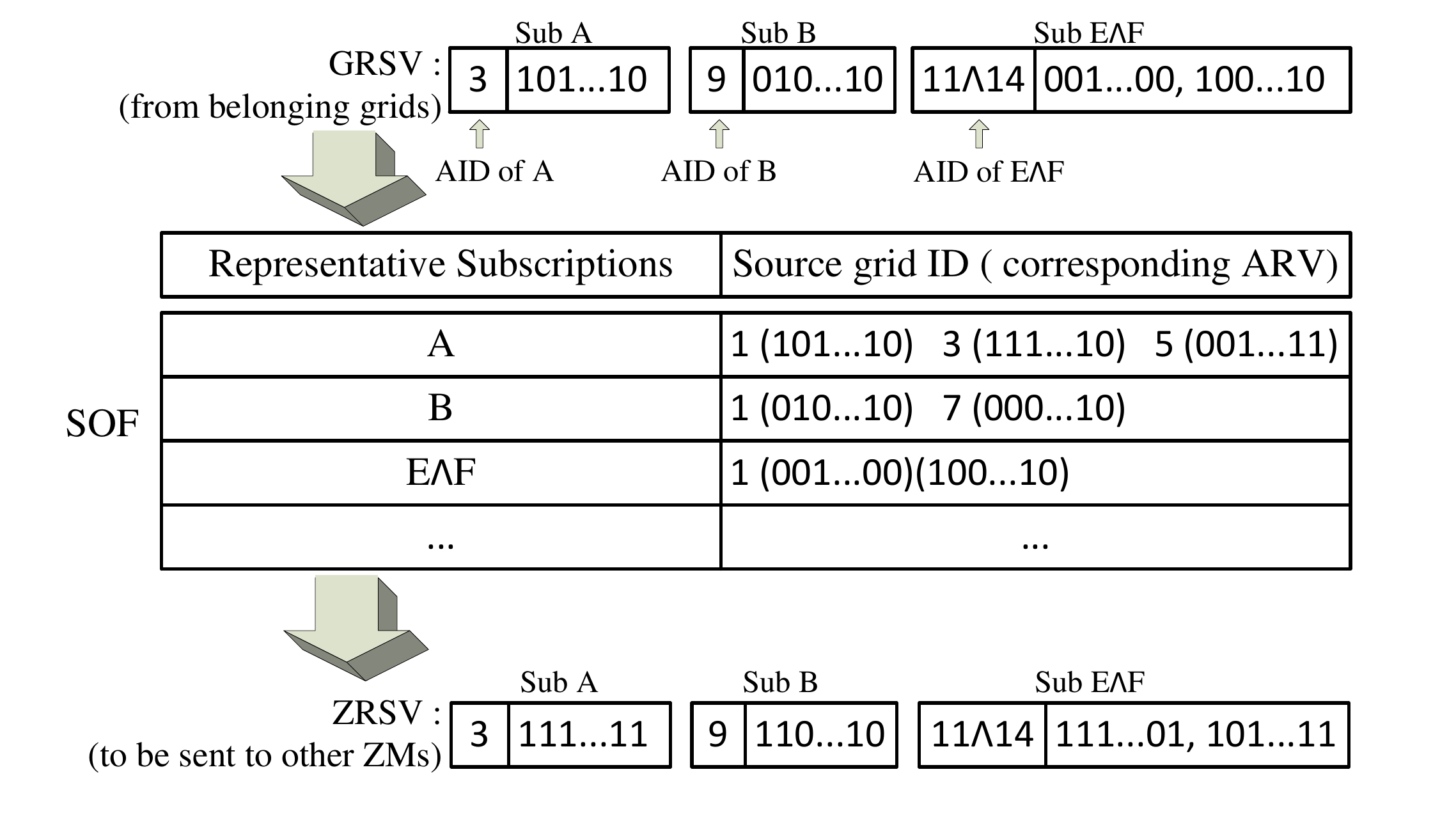}
% where an .eps filename suffix will be assumed under latex,
% and a .pdf suffix will be assumed for pdflatex; or what has been declared
% via \DeclareGraphicsExtensions.
\caption{The ZM converts the GRSVs received from belonging grids into SOF, then converts it into ZRSV by summary tree scheme.}
\label{fig:SOF2ZRSV}
%\vspace{-0.1in}
\end{figure}
Each zone manager maintains a subscription origin form (SOF) generated based on the GRSVs sent by grids with subscriptions within its zone, as shown in Figure~\ref{fig:SOF2ZRSV}. The representative subscriptions from the grids will again be aggregated through the summary tree scheme similar to that at the grid level. We cascade each subscription of the resulting representative set to form a long vector - Zone Representative Set Vector (ZRSV). The ZRSVs are exchanged among ZMs to guide the publication distributions. The SOF will be updated if there is a GRSV update, but similar to the grid level aggregation, an individual update in SOF may not lead to ZRSV change. The aggregation helps to reduce the message distribution and simplify the information matching process, which is more critical for dynamic wireless networks.
%This will significantly reduce the message overhead in a dynamic wireless network.
The ZRSV only needs to be distributed to relevant zones upon changes, after a given period, or when the zone receives unwanted traffic. Each ZM maintains the ZRSVs received from neighboring zones and zones interested in its publications (learned from previous successful match processes) to guide the distribution of published data.

\subsection{\textbf{Match a Publication over Subscriptions}}
When a node generates a publication, it will send the data along with the publication ARVs describing the data to its GM. GM will perform a match within its grid by comparing the publication ARVs with its representative Sub set, i.e., the roots with summary ranges of the summary forest, using the matching rule defined in Section~\ref{Section:ARVmatching}. If a root is matched, each of its tree node is further examined to precisely find the subscribers. The data will be forwarded to the identified subscribers through on-demand stateless geographic multicast scheme~\cite{multicastTOC}. No matter local matches are found or not, GM will forward the data along with the P-ARVs and the grid ID to the zone manager.

The ZM will match the P-ARVs against its SOF, to decide which grids within the zone to forward the data to for further matching at GM level. It also matches against all ZRSVs for other zones it maintains. The data along with the publication P-ARVs and the zone ID will be multicasted towards the centers of the zones that match this publication. Once the data reach a target zone, they will be forwarded to the ZM which will match the Pub with each item of the SOF. The data will be multicast to the matched grids, where the GMs will again multicast the data finally to the matched subscribers.

As mentioned earlier, each ZM only actively maintains the ZRSVs of its neighbors. However, other zones may also have subscribers to its publication.
%If the publication cannot match any of the ZRSVs, the ZM will record the publication ARV and temporarily store the data or store the data in a storage server in its zone.
If the publication ARVs associated with a publisher are seen by the ZM the first time or after a given time period since its last global distribution, the ARVs will be multicasted to all zones to inform them the existence of new publications. A zone $x$ with the matched ZRSV will send to the ZM its ZRSV, which will be maintained by the ZM along with other ZRSVs. ZM will multicast publication data to the zones with matched ZRSVs. A zone will update its ZRSV to the publication zone following the ZRSV update rules described earlier. A ZRSV will be removed if there are no data match with it for a predefined timeout period. To further reduce the overhead, for a large system, the period of sending the publication ARVs to farther-away zones can be made larger as generally the information has location constraints. In addition, a zone without any data matched with some subscriptions could also actively search for publishers by broadcast a query message within certain range or query the ZMs within certain zone-hop distance.
%These options depend on the information types and characteristics, and as mentioned earlier, developing more efficient global distribution scheme is not the focus of this paper.
%Upon receiving a publication data, the ZM will also cache the data in a Pub_Cache node.
\subsection{\textbf{Publication Caching and Match}}
Publications may not match any subscription in a single attempt, and a subscriber may want to retrieve earlier published data. Conventional studies generally assume publications always get matched; if not, the unmatched publications are simply discarded. This would waste the system resources that haven been used in generating, matching and distributing these published data, and also cannot meet the users' urgent needs for previously published data if discarded.
In this work, we introduce publication caching to facilitate bidirectional matching which also supports matching a subscription over cached publications. A zone manager receiving a publication will cache the data at the ZM or designated storage server for a predefined duration, and records the ARVs of this pub along with its source node's ZID and GID. In case that the caching space is running out, data with least matching-hit records will be removed.

A ZM holds SOFs of its own zone and ZRSVs of other relevant zones. Upon the update of the SOF or ZRSV,
%indicating subscription changes in the system,
the ZM will compare the changed SOF or ZRSV with the ARVs of the cached publication so that the matched subscribers can get the interested data right away.

\section{Simulation and Performance Evaluation}
\label{Section:Simulation}
We implement BRVST using NS2.34. The focus of BRVST is on information content matching and forwarding mechanisms, and the underlying routing scheme follows SOGR~\cite{SOGR} and RSGM~\cite{multicastTOC} for on-demand robust unicast and multicast respectively. 400 nodes are randomly distributed initially in a network region of size 1000m x 1000m to reflect the real-world mobile user density. In our default setting, the network is divided into 4 equal zones with 4 equal grids inside each. These numbers will vary when studying the impact of grid size on system performances.
%These numbers will vary when studying Only in the simulation for studying grid size impact on system performances, this setting of 16 grids might change according to different grid sizes picked given the size of the evaluation region is fixed.
The node movement follows the improved Random Waypoint model~\cite{RandomWaypointModel}. All the nodes including the autonomously elected GM and ZM could move following the model. The wireless channel propagation model is set to be TwoRayGround, and 802.11a is adopted as the MAC protocol with an average transmission range of 80m. Publications and subscriptions are generated by randomly selected nodes. Each publication or subscription has one to three attributes, which are randomly selected from a predefined set of 15. The range of an attribute is also randomly generated within a predefined range limit based on the attribute type. If not otherwise specified, the average node moving speed is set to 5 m/s, the Pub and Sub generation rates are both set to 200/minute, and the accuracy threshold $\alpha$ is set to 90$\%$.

%so that when three different implementations are compared their management overhead influences in forms of traffic volume on the timing performance are under a fair and consistent criteria.

There is very limited number of studies closely related to ours. For performance references, we select two existing Pub/Sub schemes, DRIP and TAMA, that are partly comparable to our work. DRIP~\cite{DRIP} (INFOCOM'08) is proposed for wireless networks which group nodes into Voronoi regions to manage the network, while BRVST introduces geographic zones to facilitate management and information distribution. TAMA~\cite{TAMA} (ICDCS'11) is a middleware for content matching, but is not specified for wireless networks. To be fair, we compare the impact of node mobility on the matching time for DRIP and BRVST in wireless environment, without including TAMA. The number of Voronoi regions for DRIP is also set to 16 under the same region area and node density. Since TAMA also considers using attribute range to describe contents, we compare it with BRVST on the false positive rate. The management overhead involved for storing and transmitting publication and subscribe filters are compared among all three schemes.

%For a given network area, the grid size impacts the total number of grids thus the number of nodes and messages per grid. This in turn affects the message storage, management traffic overhead, as well as the matching time. We study these effects in Fig.~\ref{fig:MobilityAndGridSize}-(b) by testing several discrete choices of grid size such that the corresponding number of grids needed to cover the study area varies from as abundant as 64 to the extreme case of only 1.

%In most of the simulations, key measurements are examined at varying message load, including the generation frequency of both publications and subscriptions, which reflects the most dominant characteristic of information network, and is a factor as a result of the comprehensive impact of node density and network size.

\subsection{\textbf{Matching Time}}
\label{SimulationMatchingTime}
%First DRIP and BRVST are compared under a scenario where all the nodes are uniformly distributed across the network with the same average moving speed.

It is equally important for both the information provider and consumer to be served as fast as possible, so we evaluate the time for an emergent publication and an emergent subscription to get matched separately.
%\vspace{-0.1in}
\begin{figure}[h]
\centering
\includegraphics[width=3.5in]{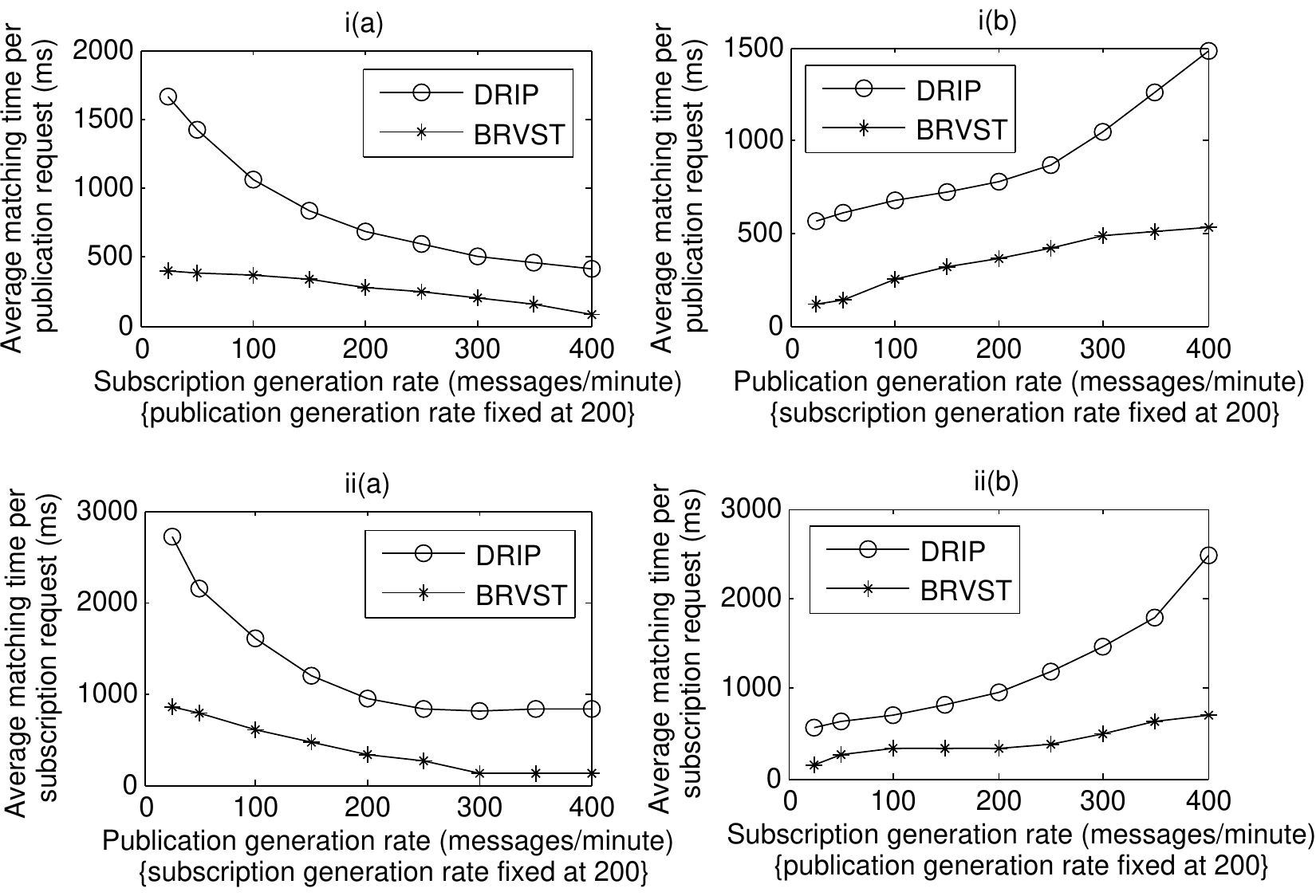}
% where an .eps filename suffix will be assumed under latex,
% and a .pdf suffix will be assumed for pdflatex; or what has been declared
% via \DeclareGraphicsExtensions.
\caption{i(a) Matching time per Pub request as Sub rate increases; i(b) Matching time per Pub request as Pub rate increases; ii(a) Matching time per Sub request as Pub rate increases; ii(b) Matching time per Sub request as Sub rate increases.}
\label{fig:MatchingTime}
%\vspace{-0.12in}
\end{figure}

For each newly published event, we evaluate the average time taken to match it with the subscribers. We allow publication to be matched with a later generated subscription and vice versa, so the delay is also affected by the subscription and publication generating frequency, as shown in Figure~\ref{fig:MatchingTime}-i. In Figure~\ref{fig:MatchingTime}-i(a) the publications rate is fixed at 200/min, while the subscription rate is varied. In Figure~\ref{fig:MatchingTime}-i(b), the subscription rate is fixed at 200/min, while the publication rate is varied. Similarly, we evaluate the average time duration for a newly generated subscription to match the publication in Figure~\ref{fig:MatchingTime}-ii(a) and (b), with the subscription and publication rate fixed at 200/min respectively.

 We can observe that BRVST has a much shorter average matching time as compared to DRIP under all test scenarios. A publication (or subscription) request has a shorter time to be matched when there is a higher subscription (or publication) rate as shown in Figures~\ref{fig:MatchingTime}-i(a) and ii(a). The reduction of matching time reaches a limit, beyond which the matching time may slightly increase as a result of higher processing overhead.

%  the publications(or subscriptions) are denser than the subscriptions(or the publications), the subscriptions(or publication) quickly get matched. As the subscription rate keeps increasing, the average matching time for a publication keeps shrinking, simply because there are more and more choices for the publications to try, enlarging the chance of match, and the curves in i(a) reflect this trend. ii(a) shows the same trend for average matching time of a subscription. However, the matching time cannot be continuously shortened by increasing the density of its counterpart. As shown in ii(a), there is a lower limit of matching time, after which, the matching time rises again due to the processing burden of increasing message density.

On the contrary, as the publication (or subscription) rate becomes larger, the time to match a publication (or subscription) increases as a result of competitions, which deteriorate the average matching time, as shown in Figures~\ref{fig:MatchingTime}-i(b) and ii(b). As DRIP involves network-wide broadcast to establish and maintain Voronoi regions, the matching time increases exponentially, while BRVST has only a sub-linear increasing time, which indicates its better scalability to system load.
%\vspace{-0.1in}
\begin{figure}[h]
\centering
\includegraphics[width=3.5in]{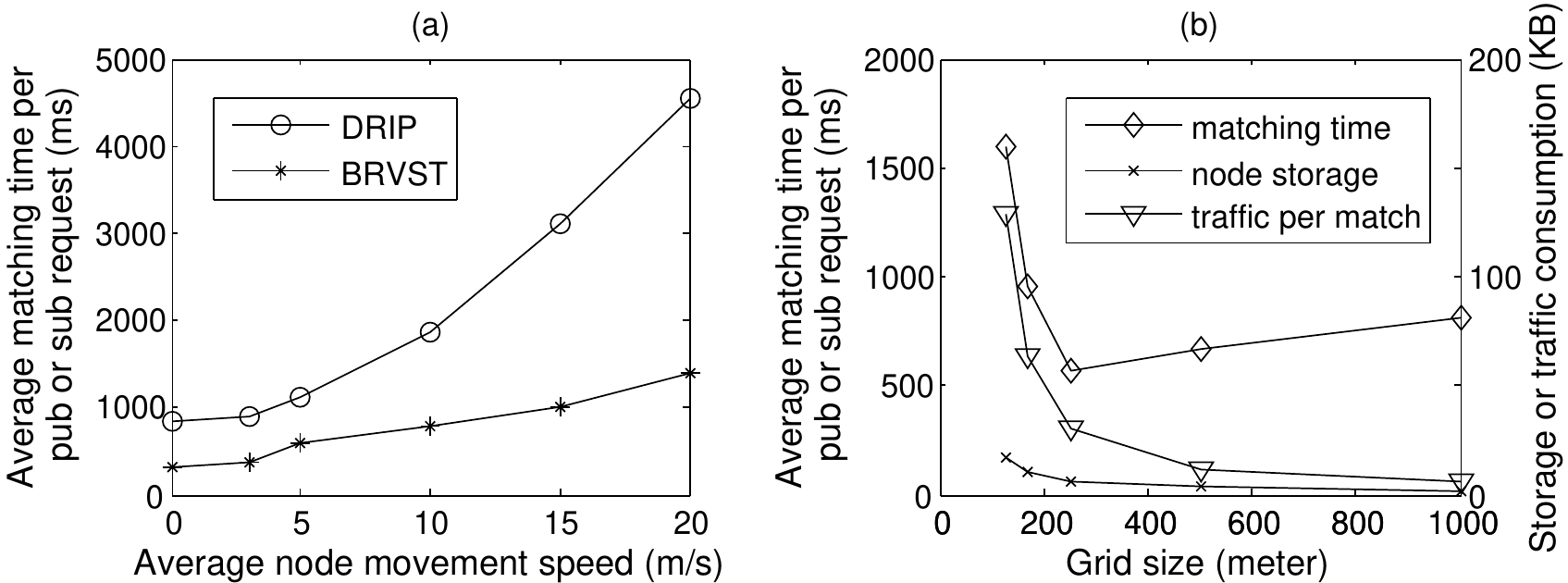}
% where an .eps filename suffix will be assumed under latex,
% and a .pdf suffix will be assumed for pdflatex; or what has been declared
% via \DeclareGraphicsExtensions.
\caption{(a)Mobility impact on average matching time; (b)Grid size impact on BRVST's average matching time per message, system average node storage consumption and traffic volume incurred per match. The setting of grid size variation corresponds to the number of grids varying from 64 downto 1.}
\label{fig:MobilityAndGridSize}
%\vspace{-0.12in}
\end{figure}

Figure~\ref{fig:MobilityAndGridSize}-(a) tests and compares the reliability of BRVST and DRIP in terms of matching time performance under high node mobility, with the average node speed varying from 0 to 20m/s. The average matching time per message (including either the publication match or subscription match) of DRIP increases significantly as a result of its broadcast-based management overhead. The delay becomes more severe when the average moving speed is higher than 10m/s, where nodes could move across regions within the average matching duration. Based on light-weight virtual management infrastructure, BRVST has much more stable performance in the mobility case.

%We study these effects by varying the size of grid such that the corresponding number of grids needed to cover the study area varies from as abundant as 64 to the extreme case of only 1 as shown in Fig.~\ref{fig:MobilityAndGridSize}-(b).

In Figure~\ref{fig:MobilityAndGridSize}-(b), the matching time is seen to first reduce with grid size and then increase.
 As the grid size increases, the number of grids decreases so does the number of zones, while the number of nodes in a grid increases. In a larger grid, messages are more likely to get matched within the grid or zone, and there are fewer other zones to check with. However when the grid size gets too large, messages need to interact over longer distance with GMs and ZMs. In addition, a large number of nodes also result in more filters in a grid which incurs a longer matching time.

\subsection{\textbf{System Maintenance Overhead}}
\label{SimulationStorageTraffic}
\begin{figure}[h]
\centering
\includegraphics[width=3.5in]{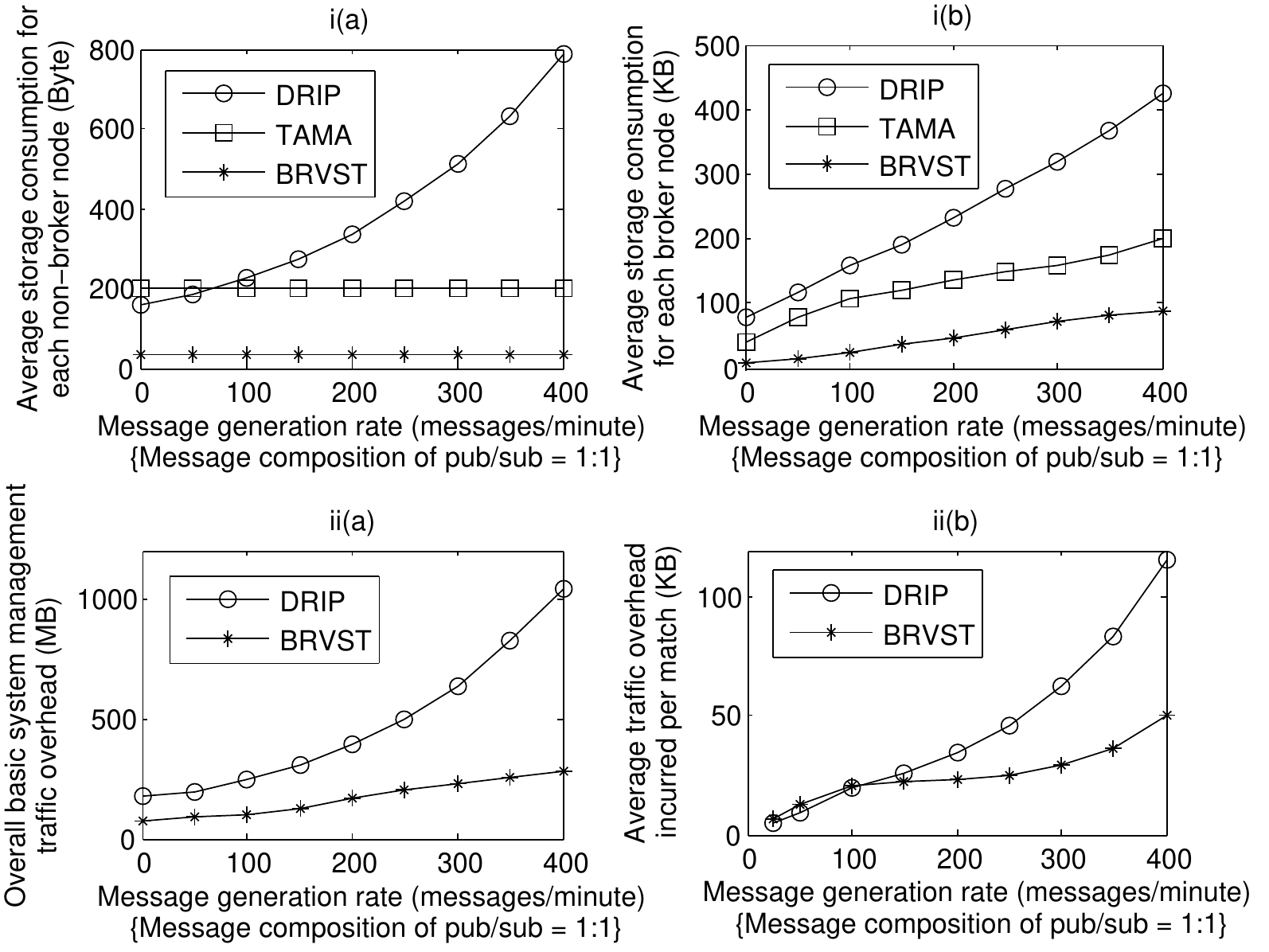}
% where an .eps filename suffix will be assumed under latex,
% and a .pdf suffix will be assumed for pdflatex; or what has been declared
% via \DeclareGraphicsExtensions.
\caption{Storage consumption for i(a)non-broker node; i(b)broker node; ii(a)Basic system traffic overhead; ii(b)Traffic overhead incurred per match.}
\label{fig:StorageTrafficOverhead}
%\vspace{-0.12in}
\end{figure}

 We compare the overhead for storing and transmitting management messages at broker nodes and regular network nodes respectively. In Figure~\ref{fig:StorageTrafficOverhead}, the publication and subscription rates increase at the same speed.

In Figure~\ref{fig:StorageTrafficOverhead}-i(a), TAMA and BRVST both have lower storage overhead at regular nodes, as these nodes do not store publication and subscription information. Specifically, BRVST only requires each node to keep a few ID numbers which are very small in volume. With the need of storing a delay list of brokers and neighboring information, DRIP has much higher storage overhead, and the overhead increases quickly with the load.

In Figure~\ref{fig:StorageTrafficOverhead}-i(b), the storage overhead at brokers for all three schemes increase linearly with the load. DRIP has a much higher increasing rate with its need of maintaining information of both non-broker nodes and other brokers, as well as the subscriptions and publications of all the nodes in the network. Both TAMA and BRVST exploit range-based content representation to reduce the storage space. BRVST exploits space efficient aggregate scheme, so its storage space is 60\% lower than that of TAMA.
%\vspace{-0.1in}

We compare DRIP and BRVST on the overhead for transmission of management messages. In Figure~\ref{fig:StorageTrafficOverhead}-ii(a), the overhead of DRIP increases exponentially due to its inefficient broadcast mechanism. BRVST does not requirement significant overhead to maintain its zone and grid infrastructure, and only sends highly aggregated publish or subscribe information, thus it has a much lower transmission overhead.

In Figure~\ref{fig:StorageTrafficOverhead}-ii(b), when the message rate is low, BRVST and DRIP have similar matching overhead. At a higher load, however, the overhead of DRIP increases exponentially, while the overhead of BRVST is compensated as each publication can match multiple subscriptions with its aggregate subscription mechanism.

In Figure~\ref{fig:MobilityAndGridSize}-(b), as grid size increases, both the average node storage space and the traffic volume incurred for each match reduce. With a larger grid size, nodes are less likely to move out of the grid, thus the overhead associated with grid change will be lower. A larger grid also allows better information aggregation, thus reducing the matching traffic.

\subsection{\textbf{False Positive Rate}}
As TAMA and BRVST represent contents with certain range granularity to reduce complexity, it would also introduce some false positive rate and forwards some unwanted traffic to nodes.
%With its granularity to be flexibly controlled and subject to an accuracy requirement, BRVST has much

\begin{figure}[h]
\centering
\includegraphics[width=3.5in]{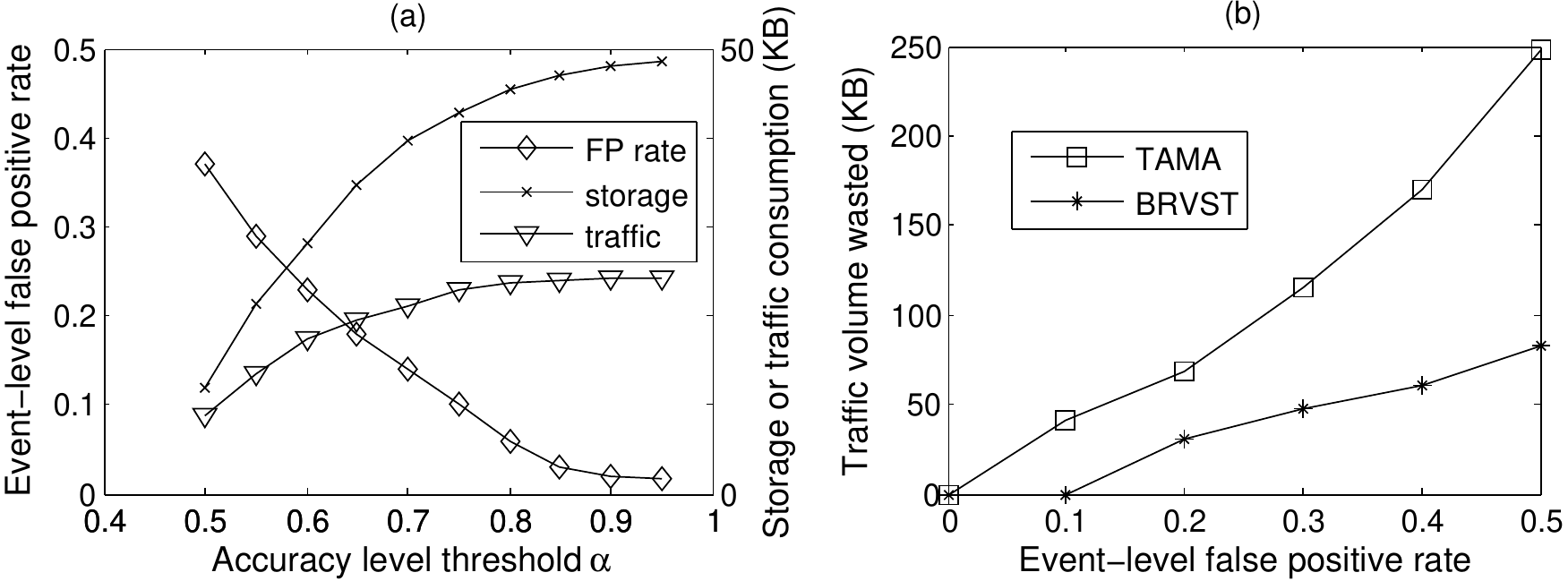}
% where an .eps filename suffix will be assumed under latex,
% and a .pdf suffix will be assumed for pdflatex; or what has been declared
% via \DeclareGraphicsExtensions.
\caption{(a)Threshold $\alpha$ impact on BRVST's event-level false positive rate, average broker storage consumption and traffic volume incurred per match; (b)Traffic volume wasted due to false positive match for BRVST and TAMA.}
\label{fig:ThresholdAndFalsePositive}
%\vspace{-0.12in}
\end{figure}

Figure~\ref{fig:ThresholdAndFalsePositive}-(a) shows that the false positive rate is inversely proportional to the accuracy threshold $\alpha$, and approximately bounded by $1- \alpha$. There is an obvious tradeoff between the accuracy level of representing information and the length of the ARV vector thus the overhead. The higher the accuracy level, the more storage and traffic volume incurred. Figure~\ref{fig:ThresholdAndFalsePositive}-(b) shows as the false positive rate rises, the traffic overhead of both system increases. However at the same false positive rate, BRVST would waste much less traffic than TAMA due to its efficient ARV representation.

 %compensation of reducing message length.}
% The finer we want the range to be represented, the more bits we need to use, and the lower the false positive rate will be. Our simulations under various Pub/Sub density cases reveal the relevance between the false positive rate and the system management overhead for both systems. In BRVST, as we flexibly modulate the extent of accuracy in range expression, we may sacrifice the false positive rate, but in turn it saves the system management messages simply because the vector length of every publication and subscription is shorter now. Figure~\ref{fig:MobilityFalsePositive}-(b) shows as false positive rate rises, both system experience traffic overhead increase. However at the same false positive rate, BRVST has much lower traffic overhead than TAMA, due to its compensation of reducing message length.}

%%\vspace{-0.1in}

%  Favored from the properties mentioned above, when the false positive rate to individual attribute is small, e.g. below 0.1, BRVST has no actual false positive for Pub/Sub matching, thus no incurred traffic is wasted. As the individual attribute false positive rate increases, the real false matchings start happening which cause growing traffic waste. The figure shows BRVST is more reluctant to false positive than TAMA as the false positive rate aggravates.

%\subsection{\textbf{Grid and Zone Size}}
%\note{The size of grid affects the system performance in a way that is equivalent to varying the number of nodes managed by each grid manager, which is reflected essentially in the information density, as discussed in Figure~\ref{fig:MatchingTime}and~\ref{fig:StorageTrafficOverhead}. With larger grid size, there are more nodes inside and less chance of moving out for each node. More nodes better facilitate the information aggregation efficiency that reduces overhead, but also suffers longer distance or more hops which results in longer matching time, as well as larger load in bottom level Pub/Sub matching. Composed by grids, the zone has the similar influence of its size.}

\section{Conclusion}
\label{Section:Conclusion}
In this paper, we present BRVST, an information content matching and forwarding engine in wireless network, which supports maximum flexibility in the expression of information content.
% without need of maintaining any topology control structures.
The most valuable contributions of BRVST are its introduction of a novel attribute range vector that can accurately represent information content with extreme efficiency both in space and computationally, and the summary tree concept that enables effective extraction and aggregation of information. All these proposed structures help significantly reduce storage and communication consumption as well as computation overhead, and ensure stable performance. Extensive simulations demonstrate that BRVST is reliable and scalable in large and dynamic wireless network conditions even under very high information load.

\bibliographystyle{IEEEtran}
% argument is your BibTeX string definitions and bibliography database(s)
\bibliography{myref}

% Generated by IEEEtran.bst, version: 1.12 (2007/01/11)
\begin{thebibliography}{10}
\providecommand{\url}[1]{#1}
\csname url@samestyle\endcsname
\providecommand{\newblock}{\relax}
\providecommand{\bibinfo}[2]{#2}
\providecommand{\BIBentrySTDinterwordspacing}{\spaceskip=0pt\relax}
\providecommand{\BIBentryALTinterwordstretchfactor}{4}
\providecommand{\BIBentryALTinterwordspacing}{\spaceskip=\fontdimen2\font plus
\BIBentryALTinterwordstretchfactor\fontdimen3\font minus
  \fontdimen4\font\relax}
\providecommand{\BIBforeignlanguage}[2]{{%
\expandafter\ifx\csname l@#1\endcsname\relax
\typeout{** WARNING: IEEEtran.bst: No hyphenation pattern has been}%
\typeout{** loaded for the language `#1'. Using the pattern for}%
\typeout{** the default language instead.}%
\else
\language=\csname l@#1\endcsname
\fi
#2}}
\providecommand{\BIBdecl}{\relax}
\BIBdecl

\bibitem{gryphon}
R.~E. Strom, G.~Banavar, T.~D. Chandra, M.~Kaplan, K.~Miller, B.~Mukherjee,
  D.~C. Sturman, and M.~Ward, ``Gryphon: An information flow based approach to
  message brokering,'' \emph{CoRR}, 1998.

\bibitem{SIENA}
A.~Carzaniga, D.~S. Rosenblum, and A.~L. Wolf, ``Design and evaluation of a
  wide-area event notification service,'' \emph{ACM Trans. Comput. Syst.},
  vol.~19, pp. 332--383, August 2001.

\bibitem{DRIP}
Q.~Yuan and J.~Wu, ``Drip: A dynamic voronoi regions-based publish/subscribe
  protocol in mobile networks,'' in \emph{INFOCOM 2008}, april 2008, pp. 2110
  --2118.

\bibitem{GeoRendezvous}
N.~Carvalho, F.~Araujo, and L.~Rodrigues, ``Reducing latency in
  rendezvous-based publish-subscribe systems for wireless ad hoc networks,''
  ser. ICDCSW '06.\hskip 1em plus 0.5em minus 0.4em\relax IEEE Computer
  Society, 2006.

\bibitem{GeographicalContentPubSub}
J.~Mocito, J.~A. Briones-Garc\'{\i}a, B.~Koldehofe, H.~Miranda, and
  L.~Rodrigues, ``Geographical distribution of subscriptions for content-based
  publish/subscribe in manets,'' in \emph{Proceedings of the ACM/IFIP/USENIX
  Middleware'08}.\hskip 1em plus 0.5em minus 0.4em\relax ACM, 2008, pp.
  102--103.

\bibitem{R-Tree}
A.~Guttman, ``R-trees: a dynamic index structure for spatial searching,'' ser.
  SIGMOD '84, 1984, pp. 47--57.

\bibitem{BloomfilterRange}
Y.~Hua, D.~Feng, and T.~Xie, ``Multi-dimensional range query for data
  management using bloom filters,'' ser. CLUSTER '07, 2007, pp. 428--433.

\bibitem{TAMA}
Y.~Zhao and J.~Wu, ``Towards approximate event processing in a large-scale
  content-based network,'' ser. ICDCS '11, pp. 790--799.

\bibitem{TreePubSub}
G.~Picco, G.~Cugola, and A.~Murphy, ``Efficient content-based event dispatching
  in the presence of topological reconfiguration,'' in \emph{ICDCS 2003}, may
  2003, pp. 234 -- 243.

\bibitem{ContentPubSubModel}
Z.~Jerzak and C.~Fetzer, ``Bloom filter based routing for content-based
  publish/subscribe,'' in \emph{Proceedings of the second international
  conference on Distributed event-based systems}, ser. DEBS '08, pp. 71--81.

\bibitem{multicastTOC}
X.~Xiang, X.~Wang, and Y.~Yang, ``Stateless multicasting in mobile ad hoc
  networks,'' \emph{IEEE Transactions on Computers}, vol.~59, no.~8, pp. 1076
  --1090, aug. 2010.

\bibitem{SOGR}
X.~Xiang, X.~Wang, and Z.~Zhou, ``Self-adaptive on-demand geographic routing
  for mobile ad hoc networks,'' \emph{IEEE Transactions on Mobile Computing},
  vol.~1, p.~99, 2011.

\bibitem{RandomWaypointModel}
W.~Navidi and T.~Camp, ``Stationary distributions for the random waypoint
  mobility model,'' \emph{IEEE Transactions on Mobile Computing}, vol.~3,
  no.~1, pp. 99--108, 2004.

\end{thebibliography}
%

%\balancecolumns
% that's all folks
\end{document}